\documentclass[11pt,twoside]{article}

\usepackage{asp2004}
\usepackage{epsf}
\usepackage{epsfig}
\usepackage{lscape}

\begin{document}
\title{Characteristics of the Galactic magneto-ionized ISM from
Faraday rotation}

\author{Marijke Haverkorn}
\affil{Jansky fellow, National Radio Astronomy Observatory}
\affil{Astronomy Department University of California at Berkeley, 601
Campbell Hall, Berkeley CA 94720, USA}

\begin{abstract}
Faraday rotation measurements of polarized extragalactic sources probe
the Galactic magnetized, ionized interstellar medium. Rotation
measures of these sources behind the inner Galactic plane are used to
explore characteristics of the structure in the spiral arms and in
interarm regions. Structure in the spiral arms has a characteristic
outer scale of a few parsecs only, whereas interarm regions typically
show structure up to scales of hundreds of parsecs. The data indicate
that in the spiral arms, the random component of the magnetic field
dominates over the regular field, but in the interarm regions the
random and regular field components may be comparable, and a few times
weaker than the random magnetic field in the spiral arms.
\end{abstract}

\section{Introduction}

The discussion whether tiny scale structure in atomic, molecular or
ionized form is part of a power law power spectrum or is made up of
overdense discrete structures is ongoing (see e.g.\ Heiles, Mason,
Deshpande, this volume). Therefore, in addition to the study of the
tiny structures themselves, studies to determine the power spectra of
the neutral and ionized medium are relevant to have a framework in
which the existence of tiny scale structure can be tested.

The velocity and density power spectra of neutral gas are extensively
probed as part of molecular cloud evolution and star formation studies
\citep{es04}. For the warm ionized component of the interstellar
medium (ISM), electron density studies mostly show Kolmogorov spectra
\citep{ars95}. The ionized gas dynamics and structure are expected to
be heavily influenced by magnetic fields threading the medium, the
characteristics of which are still very uncertain.

In this paper, we discuss fluctuations in the warm magneto-ionized
medium probed by way of Faraday rotation, to estimate typical scales of
structure and magnetic field strengths in spiral arms and in interarm
regions.

\section{Data analysis of polarized extragalactic point sources}

Polarized radiation from extragalactic point sources is altered by
Faraday rotation when propagating through the Milky Way plane, which
makes these point sources a good probe of the structure in the
Galaxy's magnetic field and electron density.\footnote{Faraday
rotation describes the rotation of the angle of linear polarization
$\phi$ due to birefringence for left and right handed circular
polarization in a magnetized, ionized medium. Faraday rotation is
wavelength dependent: $\phi \propto\mbox{RM}\,\lambda^2$, where rotation
measure RM is RM~$=0.81\int n_e[\mbox{cm}^{-3}] \, B_{\parallel}[\mu
\mbox{G}] \, dl[\mbox{pc}]$, with $n_e$ thermal electron density,
$B_{\parallel}$ magnetic field strength parallel to the line of sight
and $dl$ path length.}

However, as these sources are irregularly spaced on the sky,
performing a Fourier transform to obtain the typical scales of
structure introduces artifacts. Instead, the second order structure
function (SF) of rotation measure RM can be used, which is defined as
$D_{\mbox{RM}}(dr) = \langle
(\mbox{RM}(r)-\mbox{RM}(r+dr))^2\rangle_r$, where $dr$ is the
separation of two sources on the sky, and $\langle\rangle_r$ is an
average over every position $r$ which contains a source with a
neighboring source in a bin around $dr$ away. For a power law power
spectrum in RM, the SF will be a power law with a certain smallest
dissipation scale $l_d$ which is much smaller than the scales probed
in this paper, and an outer scale $l_0$ which is the maximum scale
found in the turbulence, believed to be the dominant scale of energy
input.

The data we use are from the Southern Galactic Plane Survey (SGPS,
McClure-Griffiths et al.\ 2005, Haverkorn et al.\ 2006a), a neutral
hydrogen and full-po\-la\-ri\-za\-tion 1.4~GHz continuum survey of the
Galactic plane, which spans an area of $253\deg < l < 357\deg$ and
$|b| < 1.5\deg$ and contains 148 polarized sources with an unambiguous
RM measurement (Brown et al.\ 2006). The data are obtained with the
Australia Telescope Compact Array (ATCA) and the Parkes 64m
single-dish telescope, and are publicly available (ATCA data only for
polarized continuum
data)\footnote{http://www.atnf.csiro.au/research/cont/sgps/queryForm.html}.

Lines of sight through discrete structures like H~{\sc ii} regions and
supernova remnants are biased due to the large electron density and
possibly magnetic field, which increases $|$RM$|$ in this direction
\citep{mwk03}. Therefore, we have used the total intensity 1.4~GHz
radio data from the ATCA combined with Parkes single-dish data to
determine which extragalactic sources have a sight line passing through
a supernova remnant or H~{\sc ii} region. The data from these sources
(about 15\% of the total sample) were then discarded.

SFs are sensitive to large-scale gradients in electron density across
the field of view. In addition, the geometrical component of the
change in magnetic field can to first order be approximated with a
linear contribution. A plane in RM is subtracted from the area in
which a SF is computed to correct for these effects.

\begin{figure}[t]
\centerline{\psfig{figure=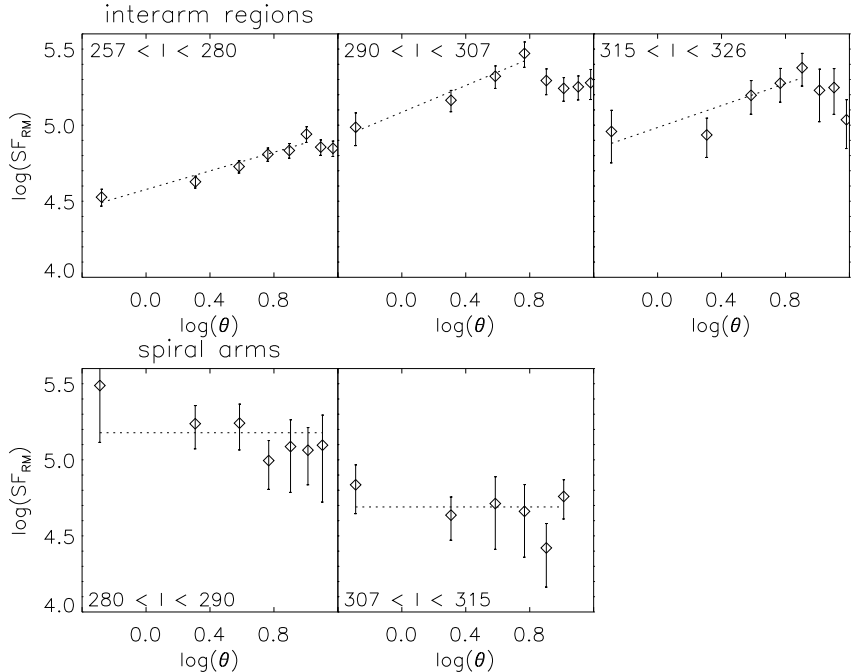,width=.85\textwidth}}
\caption{Structure functions of RM for Galactic interarm regions (top)
         and spiral arms (bottom). The dotted lines are linear fits to
         the rising parts of the SFs (top) and horizontal lines
         (bottom).}
\label{f:sf}
\end{figure}

Because the SGPS data probe the inner Galaxy, which includes a number
of spiral arms, they are well-suited to study differences in the
structure in the ISM in spiral arms and in interarm regions. The SGPS
sources are used to construct SFs for different lines of sight
preferentially through spiral arms or mostly through interarm regions,
estimated from the spiral arm positions in
\citet{cl02}. Figure~\ref{f:sf} shows the SFs in lines of sight
primarily going through interarm regions (upper panels) and in lines
of sight dominated by spiral arms (lower panels).

%***********************
%***********************

\section{Typical scales of structure}
\label{s:out}

\subsection{The turnover scale of structure functions}

The difference between the structure in RM in spiral arms and the
structure in interarm regions is obvious: the spiral arm SFs are flat,
while in interarm regions the SFs rise to a certain turnover in the
SF. The location of the turnover is interpreted as the largest angular
scale of structure in the interarm regions. With the argument that the
largest angular scales in RM are probably coming from nearby, this
outer scale corresponds to spatial scales of about 100-200~pc. For the
spiral arms we can only estimate an upper limit for the outer scale of
structure, i.e.\ the smallest scale we probe. In this way, we estimate
the outer scale of structure in the spiral arms to be smaller than
about 10~pc \citep{hgb06}.

\begin{table}
\begin{center}
\begin{tabular}{cccccccccc}
region & range in l & $r_{out}$ & $p$ & $\sigma$ & $r_{out}^d$ & $B_0$ & $C_B^2$ & $B_{ran}$ \\
& [$^{\circ}]$ & [pc] & [\%] & [rad m$^{-2}$] & [pc] & [$\mu$G] & & [$\mu$G]\\
\hline
Inter1 & 255 - 281  & 100    & 5.5 & 200 & 1 & 2.8 &  10 & 2.1\\
Carina & 281 - 292  & $<$ 17 & 1.5 & 250 & 7 & 3.5 &  18 & 2.8\\ 
Inter2 & 292 - 308  & 170    & 3.7 & 250 & 3 & 3.8 &   3 & 1.2\\
Crux   & 308 - 317  & $<$ 40 & 3.3 & 160 & 6 & 3.2 & 100 & 6.7\\
Inter3 & 317 - 327  & 220    & 2.5 & 225 & 5 & 3.7 & 100 & 6.7\\
\hline
\end{tabular}
\caption{ISM parameters for three interarm regions and the Carina and
  Crux spiral arms. The parameter $r_{out}$ is the outer scale
  determined from the turnover of the structure functions; $p$ is
  observed polarization degree, and $\sigma$ is the standard deviation
  in RM. The outer scale as determined from depolarization is given by
  $r_{out}^d$, $B_0$ is the parallel component of the regular magnetic
  field, $C_B^2$ is the amplitude of the magnetic field spectrum given
  in 10$^{-13}$ m$^{-2/3}$ $\mu$G$^2$, and $B_{ran}$ the resulting
  random magnetic field strength.\label{t:out}}
\end{center}
\end{table}

\subsection{Depolarization of point sources}

%***********************
\begin{figure}[t]
\centerline{\psfig{figure=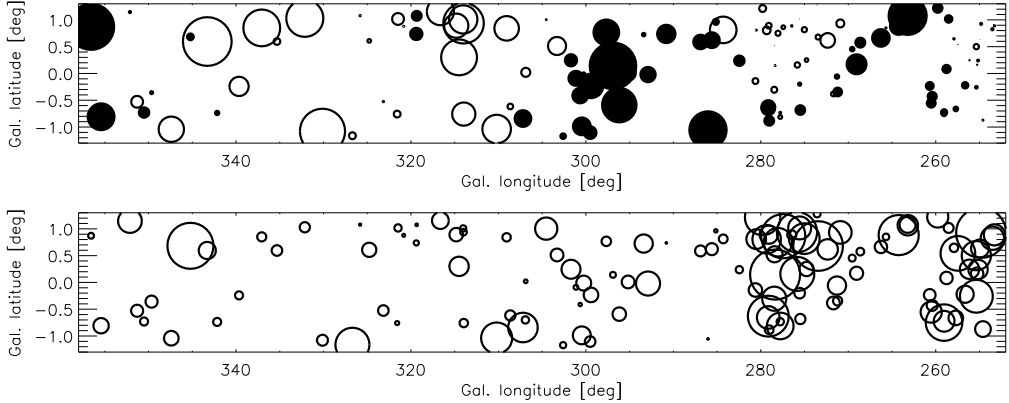,width=\textwidth}}
\caption{Top: SGPS field of view, where the circles are RMs of
         extragalactic sources. RM minimum and maximum is
         $\pm$1000~rad~m$^{-2}$. Bottom: same field and same sources,
         but the circles denote degree of polarization. Minimum degree
         of polarization is 0.4\%, the maximum 13.7\%.}
\label{f:depol}
\end{figure}
%***********************
%***********************
\begin{figure}[t]
\centerline{\psfig{figure=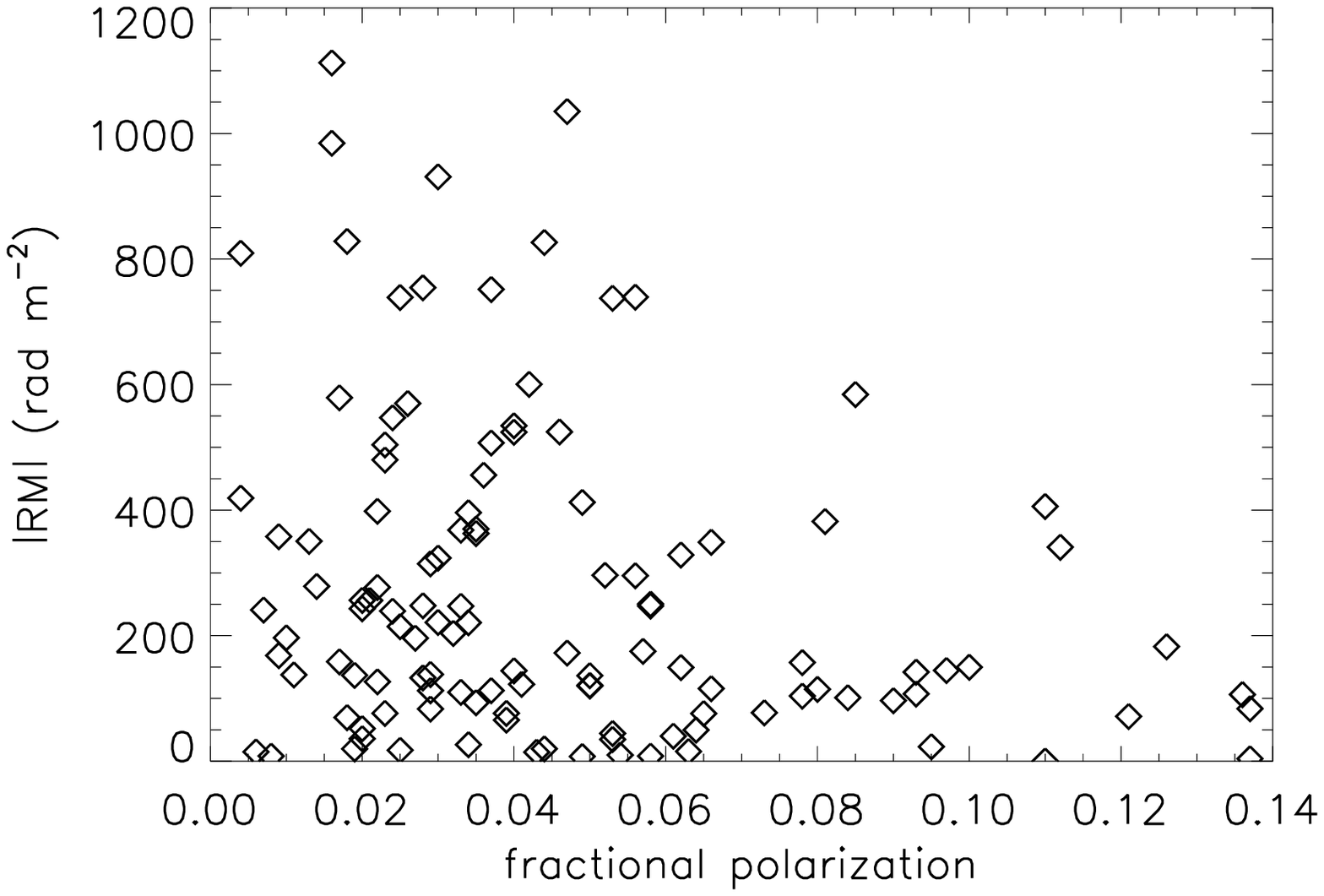,width=.55\textwidth}}
\caption{Polarization degree $p$ against $|$RM$|$ for each source.}
\label{f:rm_p}
\end{figure}
%***********************

An independent estimate of the outer scales of the structure can be
made from depolarization of extragalactic point sources.  This
depolarization is caused by variability in polarization angle on
angular scales smaller than the size of the (unresolved) source. The
variance in polarization angle within a telescope beam will decrease
the polarization degree of the source.

Variations in polarization angle causing partial depolarization are
expected to arise within any polarized extragalactic source
itself. Indeed, no source in our sample exhibits the intrinsic maximum
degree of polarization of around 70\%, but instead observed
polarization degrees are typically under 10\%. However, a Galactic
component to this depolarization has been detected as
well. Figure~\ref{f:depol} shows RMs in the upper panel and
polarization degree $p$ in the bottom panel, and a clear
anticorrelation between $|$RM$|$ and $p$ is visible especially at the
lower longitudes. This trend is also evident in Figure~\ref{f:rm_p},
which shows the degree of polarization of every source as a function
of its RM. As the scale of the structure in RM and $p$ is several
degrees, this cannot be intrinsic to the sources but instead must be
caused by the Galactic ISM.

The Galactic component of this depolarization can be estimated
assuming a power-law power spectrum of RM fluctuations, in the
approximation that the outer scale of structure $r_{out}$ is much
larger than the source size $r_{src}$ which is the case here. Adapted
from Tribble (1991), the depolarization by a power spectrum of RM
fluctuations is given by the degree of polarization $p$ as:
\begin{equation}
  \langle |\frac{p(\lambda)}{p_0}|^2\rangle \approx 1 - 4 \sigma^2 
  \lambda^4 2^{m/2} \left(\frac{r_{src}}{r_{out}}\right)^m 
  \Gamma(1+\frac{m}{2})
  \label{e:tribble}
\end{equation}
where $p_0$ is the intrinsic polarization degree of the extragalactic
source radiation when it exits the source and $\sigma$, $m$ and
$r_{out}$ are defined via the structure function
\[
  D_{\mbox{RM}}(r) = \left\{
  \begin{array}{ll} 
    2 \sigma^2 (r/r_{out})^m & \mbox{for}~r < r_{out} \\
    2 \sigma^2               & \mbox{for}~r > r_{out}
  \end{array}
\right.
\]

The average degree of polarization in the studied regions is given in
Table~\ref{t:out}. The spiral arms seem to be more depolarized than
the interarm regions, with some possible confusion closest to the
Galactic center due to superposition of arms and interarm regions
along the line of sight.

The amount of intrinsic depolarization resulting in polarization
degree $p_0$ can be estimated from the extragalactic sources observed
in and around the LMC (Gaensler et al.\ 2005) to be 10.4\%, which we
assume is the average polarization degree of point sources for which
all depolarization is intrinsic. With these assumptions, the
depolarization beyond 10.4\% is due to the variations in Galactic RM
across the face of the source, which is on average 6~arcsec (Gaensler
et al.\ 2005). This percentage is higher than the actual average
degree of polarization due to a selection of strong, highly-polarized
sources over weak, weakly polarized ones. However, as we are
interested in the relative depolarization only, this selection effect
does not influence our conclusions.

In the spiral arms it is straightforward to use Eq.~(\ref{e:tribble})
to obtain the outer scale $r^d_{out}$ needed to obtain the observed
depolarization. We assume Kolmogorov turbulence ($m = 5/3$, however,
see Section~\ref{s:slope}), and determine the value of the RM standard
deviation $\sigma$ from the SF saturation level. The distance is
chosen to be the average distance to the region probed, which has a
large error due to the large spatial extent of the gas.

For the interarm regions we observe a shallow spectrum. Assuming that
this spectrum turns over to a steeper Kolmogorov spectrum towards
small scales (see Section~\ref{s:amp}), the Kolmogorov slope on small
scales will dominate the depolarization of the point sources. Then,
using linear fits to the rising parts of the slopes, we can calculate
at which $r_{out}$ and $\sigma_{RM}$ the depolarization given by
Eq.~(\ref{e:tribble}) equals the observed depolarization in
Table~\ref{t:out}. This $r_{out}$ is the outer scale of the Kolmogorov
turbulence, i.e.\ the scale at which the Kolmogorov slope turns over
into a shallower slope. This scale is given in Table~\ref{t:out} as
the $r_{out}^d$ in the interarm regions, and is consistent with the
outer scale of Kolmogorov turbulence found in the spiral arms of a few
parsecs.

\section{Amplitude of magnetic field fluctuations}\label{s:amp}

While the turnover in the SF of RM corresponds to the outer scale of
structure, the {\it amplitude} of the SF gives information about the
magnitude of the magnetic field fluctuations in the medium.

Minter \& Spangler (1996, MS96) developed a formalism with which to
describe the SF of RM assuming power spectra in magnetic field and in
electron density fluctuations which are zero-mean, isotropic and
Gaussian.  Assuming Kolmogorov turbulence, MS96 find that the RM
structure function can be described as:
\begin{eqnarray}
  D_{RM}\!\!\!&=&\!\!\! \left[\!\right. 251.226\left[ \left({\small 
    \frac{n_0}{0.1~\mbox{cm}^{-3}}}\right)^2\! \left({\small 
    \frac{C_B^2}{10^{-13}\mbox{ m}^{-2/3}\mu\mbox{G}^2}}\right)
    \!+\! \left({\small \frac{B_0}{\mu\mbox{G}}}\right)^2\! 
    \left({\small\frac{C_n^2}{10^{-3} \mbox{ m}^{-20/3}}}\right)\right] 
    \nonumber\\
    &+&\!\!\!\!23.043\!\left({\small\frac{C_n^2}{10^{-3} \mbox{ m}^{-20/3}}}\right)\!
    \left({\small \frac{C_B^2}{10^{-13}\mbox{ m}^{-2/3}\mu\mbox{G}^2}}\right)\!
    \left({\small \frac{l_0}{\mbox{pc}}}\right)^{2/3} \!\!\left.\right] 
    \!*\!\left({\small \frac{L}{\mbox{kpc}}}\right)^{8/3}\!
    \left({\small \frac{\delta\theta}{\mbox{deg}}}\right)^{5/3}
    \nonumber
\end{eqnarray}
where $n_0$ is the mean electron density, $B_{0}$ is the mean magnetic
field strength along the line of sight, $l_0$ the outer scale of
structure, $L$ the length of the line of sight and the magnetic field
and density fluctuations are described by power laws with the same
outer scale and spectral index such that
\[
%  \left<\delta n(\mathbf{r_0})\delta n(\mathbf{r_0+r}) \right> &=&
%  \int d^3q \frac{C_n^2 \mbox{e}^{-i\mathbf{q}\cdot\mathbf{r}}}{(q_0^2+q^2)^{\alpha/2}}\\
  \left<\delta B_i(\mathbf{r_0})\delta B_i(\mathbf{r_0+r}) \right> =
  \int d^3q \frac{C_B^2 \mbox{e}^{-i\mathbf{q}\cdot\mathbf{r}}}{(q_0^2+q^2)^{\alpha/2}}
\]
and a similar expression for $\left<\delta n(\mathbf{r_0})\delta
n(\mathbf{r_0+r}) \right>$.  Assuming that on smaller scales the
magnetic field power spectrum follows the observed Kolmogorov spectrum
of electron density, the observed SFs will turn over to steeper slopes
towards smaller scales.  The constraint that the Kolmogorov SF on the
small scales and the shallower SF on larger scales must have the same
amplitude at turnover scale $l_0$ yields

\[
  D_{RM}(\delta\theta) = \left\{
  \begin{array}{ll}
    A \delta\theta^{5/3}           & \mbox{for}~\delta\theta \le l_0/L \\
    A \delta\theta^{m} (l_0/L) & \mbox{for}~\delta\theta \ge l_0/L
  \end{array}
  \right.
\]

where $m$ is the spectral index of the shallower SF. Following
this formalism, we can derive the amplitude of the magnetic field
fluctuations in the spiral arms and interarm regions, which is shown
as the dotted lines in Figure~\ref{f:sf}, while the input and output
parameters for the computation are are given in Table~\ref{t:out}. The
mean electron density was determined from the \citet{cl02} electron
density model, $B_{\parallel}$ was calculated assuming a constant
circular field with a strength $B_{reg}=4~\mu$G \citep{bbm96}, and the
outer scale of Kolmogorov turbulence $r_{out}$ is taken as 5~pc in
both arms and interarms, consistent with the rough estimates in
Table~\ref{t:out}. The path length is chosen as the distance to the
point for which 90\% of the electron density along the line of sight
is contained in the path length.

This procedure allows us to calculate the random magnetic field
coefficient $C_B^2$ and the corresponding random magnetic field
strengths of about 7~$\mu$G in the spiral arms and 2~$\mu$G in the
interarm regions, see Table~\ref{t:out}.  Caution needs to be taken
that the uncertainties in the input parameters are large so that the
magnetic field strength values are also fairly uncertain. However, the
analysis indicates that the random magnetic field component is
consistent with a constant in all interarm regions much lower than the
value in the spiral arms.

Random magnetic field strength in the spiral arms exceeding that in
the interarms has been observed in some external galaxies (e.g.\
NGC~4631, Beck \& Hoernes 1996; IC342, Krause, Hummel \& Beck 1989),
although the situation is not clear in the Milky Way. It is expected
for spiral galaxies with weak dynamos (Shukurov 1998).

\section{Speculations on the nature of the structure}
\label{s:slope}

If the above assumption is correct and the computed SFs will turn over
to a (steeper) Kolmogorov spectrum at smaller scales, several
mechanisms can be responsible for creating the shallow slopes.

Superposition of two spiral arms with similar spatial outer scales at
different distances will yield a shallow transition SF at scales just
smaller than the saturation scale. However, lines of sight with more
spiral arm superpositions (higher longitude) should give shallower
spectra, contrary to what is observed. Furthermore, other observations
in the outer Galactic plane (Sun \& Han 2004) and at higher Galactic
latitudes (Haverkorn et al.\ 2003) which find shallow slopes without
possible spiral arm superpositions argue against this explanation.

Discrete structures with internal turbulence within a turbulent
medium, such as H~{\sc ii} regions, can also explain the
observations. In this case, on small scales (i.e.\ scales smaller than
the size of the region) turbulence in the H~{\sc ii} regions would
dominate the SF, whereas on larger scales these clouds would just add
a constant 'noise' term, which makes the total slope shallower.

Shallow SF slopes can also be caused by a transition from 2D to 3D
turbulence as suggested by MS96. They invoke physical sheets of gas in
which turbulence cannot operate perpendicular to the sheets, which is
a logical choice for the region of the sky that they probe, which has
a large H~{\sc ii} region near by. For data over a significant part of
the plane, this interpretation is not likely.

A plausible option is multiple scales of energy input in the interarm
regions: for supernova-driven turbulence, the outer scale is believed
to be about 100~pc (as observed). However, if energy sources such as
stellar winds or outflows, interstellar shocks or H~{\sc ii} regions
input a significant amount of energy into the interstellar turbulence
on smaller scales (typically parsecs, Mac Low 2004), this may flatten
the SF on scales of order 1~pc to scales of order 100~pc, as observed.

Alternatively, the power spectrum of magnetic field fluctuations may
not follow the Kolmogorov scaling at all. If this is the case, the
difference between spiral arms and interarm regions may be due to the
absence and presence of a strong regular magnetic field in the arms
and interarms, respectively, causing a different spectral index
(e.g. Schekochihin et al.\ 2004), or due to a transition between
subsonic and supersonic turbulence.

\section{Summary and conclusions}
\label{s:sum}

Faraday rotation measurements of polarized extragalactic sources
behind the inner Galactic plane in the fourth quadrant are used to
study the characteristics of the magnetized, ionized interstellar
medium in the plane, in particular in the spiral arms and in interarm
regions. Structure functions show that the typical outer scale of
structure in the spiral arms is a few parsecs, whereas in the interarm
regions fluctuations up to hundreds of parsecs in size are
observed. Partial depolarization of the extragalactic sources by the
ISM is used to derive a turbulent outer scale of a few parsecs,
assuming Kolmogorov-like turbulence. From the saturation amplitudes of
the structure functions the strength of the random magnetic field
component in the spiral arms and interarm regions is
estimated. Assuming an equal regular magnetic field strength in both
arms and interarms, it is found that the random field in the arms is
about 3 times stronger than that in the interarm regions. In this
case, the random magnetic field dominates in the spiral arms, whereas
the regular and random components are similar in the interarm regions.

\acknowledgements 

The ATCA is part of the Australia Telescope, which is funded by the
Commonwealth of Australia for operation as a National Facility managed
by CSIRO. The author would like to thank Bryan Gaensler, Jo-Anne
Brown, Alexander Schekochihin, Stanislav Boldyrev, Steve Spangler, and
Joel Weisberg for stimulating discussions. M.H. acknowledges support
from the National Radio Astronomy Observatory (NRAO), which is
operated by Associated Universities Inc., under cooperative agreement
with the National Science Foundation.


\begin{thebibliography}{}
\bibitem[Armstrong et al.(1995)]{ars95} 
  Armstrong, J. W., Rickett, B. J., \& Spangler, S. R. 1995, ApJ,
  443, 209 
\bibitem[Beck et al.(1996)]{bbm96} 
  Beck, R., Brandenburg, A., Moss, D., et al. 1996, ARA\&A, 34, 155 
\bibitem[Beck \& Hoernes(1996)]{bh96} 
  Beck, R., \& Hoernes, P. 1996, Nat, 379, 47
\bibitem[Brown et al.(2006)]{bhg06} 
  Brown, J. C., Haverkorn, M., Gaensler, B. M., Taylor, A.~R., et al.
  2006, ApJL submitted
\bibitem[Cordes \& Lazio(2002)]{cl02} 
  Cordes, J.~M., \& Lazio, T.~J.~W. 2002, preprint (astro-ph/0207156)
\bibitem[Elmegreen \& Scalo(2004)]{es04}
  Elmegreen, B.~G., \& Scalo, J. 2004, ARA\&A, 42, 211
\bibitem[Gaensler et al.(2005)]{ghs05} 
  Gaensler, B.M., Haverkorn, M., Staveley-Smith, L., et al.  2005, 
  Science, 307, 1610
\bibitem[Haverkorn et al.(2006a)]{hgm06}
  Haverkorn, M., Gaensler, B.~M., McClure-Griffiths, N.~M., et
  al. 2006a, ApJ, in press
\bibitem[Haverkorn et al.(2006b)]{hgb06}
  Haverkorn, M., Gaensler, B.~M.,  Brown, J. C., Bizunok, N.,
  et al. 2006b, ApJL, 637, 33
\bibitem[Haverkorn et al.(2003)]{hkb03} 
  Haverkorn, M., Katgert, P., de Bruyn, A. G. 2003, A\&A, 403, 1045
  %SFs 
\bibitem[Krause et al.(1989)]{khb89}
  Krause, M., Hummel, E., \& Beck, R. 1989, A\&A, 217, 4
\bibitem[Mac Low(2004)]{m04}
  Mac Low, M.-M. 2004, Ap\&SS, 289, 323
\bibitem[McClure-Griffith et al.(2005)]{mdg05}
  McClure-Griffiths, N.~M., Dickey, J.~M., Gaensler, B.~M., et
  al. 2005, ApJS, 158, 178
\bibitem[Minter \& Spangler(1996)]{ms96} 
  Minter, A. H., \& Spangler, S. R. 1996, ApJ, 458, 194
\bibitem[Mitra et al.(2003)]{mwk03}
  Mitra, D., Wielebinski, R., Kramer, M., \& Jessner, A. 2003, A\&A,
  398, 993
\bibitem[Schekochihin et al.(2004)]{sct04}
  Schekochihin, A.~A., Cowley, S.~C., Taylor, S.~F., et al. 2004, ApJ,
  612, 276
\bibitem[Shukurov(1998)]{s98}
  Shukurov, A. 1998, MNRAS, 299, 21
\bibitem[Sun \& Han(2004)]{sh04} 
  Sun, X.~H, \& Han, J.~L. 2004, in The Magnetized Interstellar
  Medium, ed.\ B.~Uyan\i ker, W.~Reich, R.~Wielebinski
  (Katlenburg-Lindau: Copernicus GmbH), 25
\bibitem[Tribble(1991)]{t91}
  Tribble, P.~C. 1991, MNRAS, 250, 726

\end{thebibliography}
\end{document}